\documentclass{article}
\usepackage{amsmath,graphicx,mlspconf}
\usepackage{xcolor}

\usepackage{comment}
\usepackage{hyperref}
\usepackage{cite}
\usepackage{amssymb,amsfonts}
\usepackage{algorithmic}
\usepackage{textcomp}
\usepackage{epsf}
\usepackage{times}
\usepackage{float}
\usepackage{wasysym}
\usepackage{color}
\usepackage{stfloats}
\usepackage{subeqnarray}
\usepackage{multirow}
\usepackage{booktabs}
\usepackage{subcaption}
\usepackage{url}
\usepackage{tablefootnote}

\toappear{2026 IEEE International Workshop on Machine Learning for Signal Processing, Sep.\ 28-- Oct.\ 1, 2026, Atlanta, USA}

\title{Dynamic Clustering for Cross-Segment Permutation Alignment \\ in Long Speech Separation}
\name{Anonymous\thanks{Anonymous.}}
\address{Anonymous}

\name{Yuzhu Wang$^{1}$,
      Archontis Politis$^{1}$,
      Konstantinos Drossos$^{2}$,
      Tuomas Virtanen$^{1}$}
\address{$^{1}$Signal Processing Research Centre, Tampere University, Tampere, Finland \\
$^{2}$Nokia Technologies, Espoo, Finland
}
\begin{document}
\ninept

\maketitle

\begin{abstract}
Long speech separation typically employs a segment-separation-stitch paradigm where recordings are divided into short segments, processed independently, and stitched together. Its challenge lies in predicting cross-segment permutations. 
This paper proposes a training-free dynamic clustering approach for cross-segment permutation alignment using speaker embedding reference pools. 
The method predicts the permutation using the cosine similarity between current segment embeddings and the reference pools. 
The approach updates reference pools by retaining the most representative speaker embeddings based on their overall cosine similarity with existing references.
As a plug-and-play post-processing module compatible with existing separation models, the proposed method demonstrates superior performance compared to existing methods on dense and sparse long speech scenarios, particularly in challenging sparse scenarios with extended utterance gaps, and further shows robustness to speaker count estimation errors in unknown speaker count scenarios.
\end{abstract}

\newcommand{\cem}[1]{\textcolor{blue}{cem: #1}}
\section{Introduction}
\label{sec:introduction}
Speech separation is the task of isolating individual speaker signals from mixtures. Deep neural network (DNN) based methods have become mainstream~\cite{wang2018supervised, vincent2018audio, araki202530+}. 
Most current research focuses on separation models operating on short segments, typically ranging from a few seconds to tens of seconds~{\mbox{\cite{luo2019conv, Yu2017permutation, kolbaek2017multitalker, Wang2023TFGridNet2}}}.
However, practical applications often involve arbitrarily long recordings or continuously streamed audio such as meetings, lectures, and broadcasts.

Separating arbitrarily long recordings can be achieved by enhancing the generalization capability of separation models~\cite{luo2020dual, li2021dualmodeling, wang2025multi, wang2025attractor}, but such methods typically generalize only to recording lengths of several minutes before facing computational complexity explosion and performance degradation. A more scalable alternative involves segmenting long recordings into shorter segments during inference, performing separation on each segment independently, and then stitching the results into complete outputs~\cite{chen2020continuous, bredin2023pyannote, von2024meeting, han2021continuous}.
Theoretically, this paradigm can handle arbitrarily long recordings. 
Since separation models are typically speaker-independent and trained using permutation invariant training (PIT)~\cite{Yu2017permutation}, the assignment of speakers to outputs is arbitrary and inconsistent between segments, and the performance of such approaches is therefore dominated by the accuracy of cross-segment permutation assignment.

Existing solutions to the cross-segment permutation problem can be categorized into three types. The first approach is overlap matching, as used in continuous speech separation (CSS)~\cite{chen2020continuous}, which determines the permutation of each segment by matching the overlapping region between adjacent segments. This approach is computationally efficient and requires no speaker embeddings, but only ensures local utterance continuity within each output stream without resolving the global permutation across the full recording.
The second approach applies global clustering to speaker embeddings extracted from separated outputs~\cite{bredin2023pyannote}. The standard pipeline extracts speaker embeddings from voice activity detection (VAD)-segmented speech regions and groups them using clustering algorithms such as $k$-means, spectral clustering, or agglomerative clustering, assigning speaker labels globally across all segments. 
However, global clustering is limited to offline processing. Residual interference in separated outputs increases the risk of VAD errors that introduce unreliable embeddings, which can distort clustering results by shifting centroids or causing misassignment across speaker clusters, with the severity depending on the robustness of the chosen clustering algorithm~\cite{von2024meeting}.
The third approach maintains a speaker cache to guide permutation alignment. Static cache methods, such as speaker inventory~\cite{han2021continuous}, pre-detect single-active-speaker regions from the mixture to build a fixed speaker reference that is used without further updates. Dynamic speaker cache methods, as employed in streaming speaker diarization such as the speaker-tracing buffer (STB)~\cite{xue2021online} and more recently arrival-order speaker cache (AOSC)~\cite{medennikov2025streaming}, maintain a set of continuously updated speaker representations for streaming identity tracking, proven effective in speaker diarization. 

All three approaches are used in meeting processing systems where the primary goal is automatic speech recognition (ASR) and transcription. In such systems, cross-segment permutation consistency receives limited attention, as individual utterances are recognized independently after VAD segmentation, and utterance-level alignment is sufficient to produce accurate recognition results. 
In contrast, human listening applications such as hearing aids, multi-party call recording, and interview production require consistent speaker identity across multiple utterances. Permutation errors cause different speakers' voices to alternate within the same output channel, severely degrading the perceptual quality.
Dedicated study of cross-segment permutation alignment for separated audio is lacking, particularly for sparse speech scenarios with extended utterance gaps.

\begin{figure}[tb]
\centering
\includegraphics[width=0.95\linewidth]{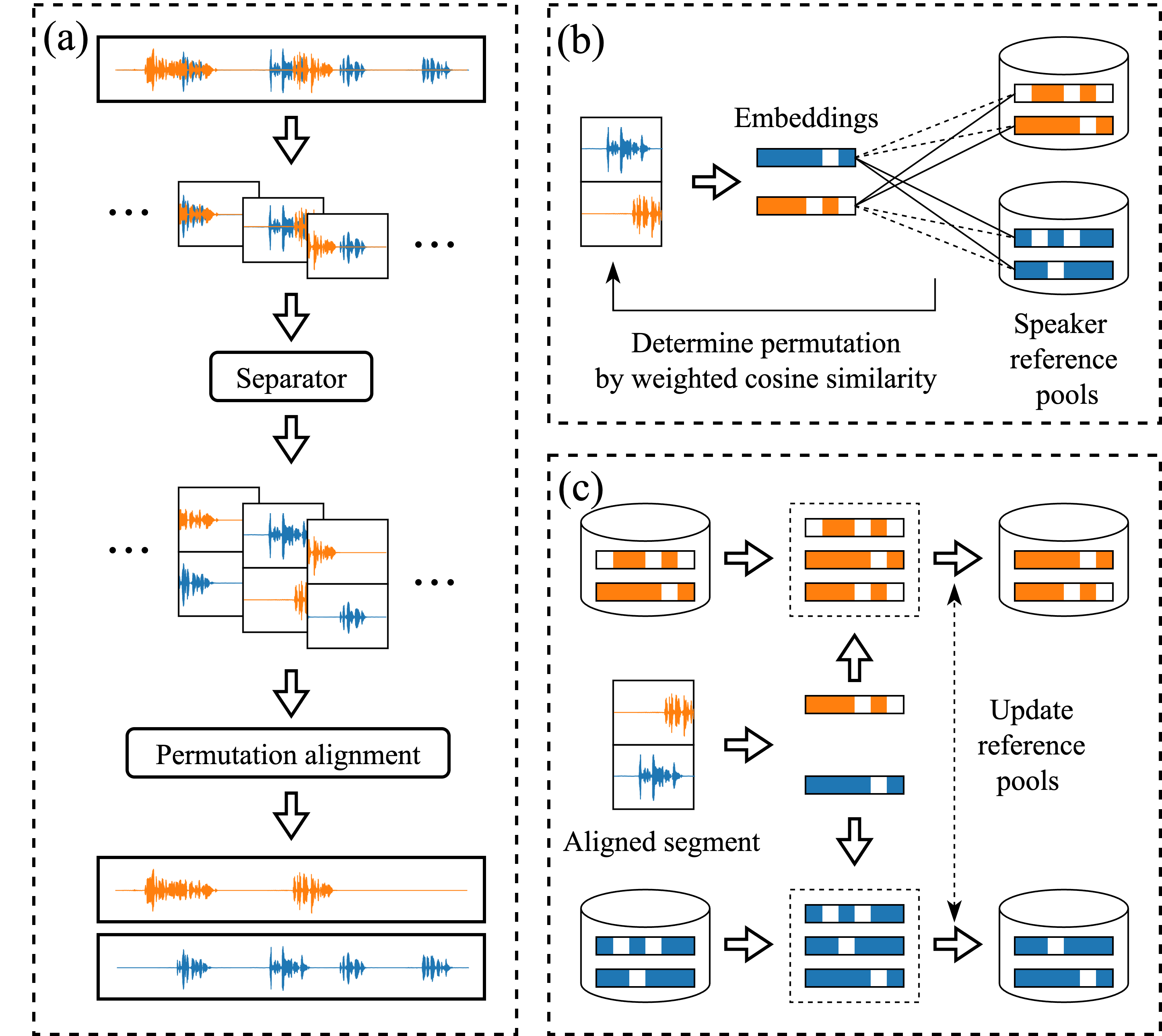}
\vspace{-1ex}
\caption{Dynamic clustering for cross-segment permutation: (a) long speech separation workflow, (b) quality-weighted permutation assignment using reference pools, (c) reference pool updates based on segment quality assessment.}
\label{fig-problem}
\vspace{-2ex}
\end{figure}

To address the cross-segment permutation problem in long speech separation, we propose a dynamic clustering approach built upon the speaker cache strategy, as shown in Fig.~\ref{fig-problem}. Similar to dynamic cache methods in streaming speaker diarization~\cite{xue2021online, medennikov2025streaming}, the proposed method maintains a reference pool for each speaker and resolves cross-segment permutations by matching current segment embeddings against cached historical representations. However, speaker diarization systems typically operate on mixture signals and output temporal speaker activity labels targeting ASR pipelines~\cite{horiguchi2022online, chen2024attention}, whereas the proposed method operates directly on separated outputs to produce speaker-consistent audio streams. Furthermore, recent dynamic cache methods rely on neural network prediction scores for cache updates~\cite{xue2021online, medennikov2025streaming}, making their generalization performance dependent on the match between training data and the target scenario. 
In contrast, since separation quality varies across segments and low-quality embeddings from poorly separated regions are unavoidable, the proposed method measures embedding reliability through inter-embedding weighted cosine similarity and retains only the most representative embeddings in the reference pool via a TopN update strategy, naturally suppressing low-quality embeddings without relying on VAD.
The proposed method requires no additional training and functions as a plug-and-play post-processing module compatible with existing separation models. Evaluation on dense and sparse long speech scenarios shows that the proposed method achieves robust performance compared to existing approaches, with particularly substantial improvements in sparse scenarios.
The proposed method also exhibits robustness to speaker count 
estimation errors, maintaining superior performance over baselines even when the speaker count is incorrectly estimated.

\vspace{-1ex}
\section{Problem Formulation}
\label{sec:problem-formulation}
Consider a long-duration mixture signal $y(t) = \sum_{c=1}^{C} x_c(t) + n(t)$ containing speech from $C$ unknown speakers and noise $n(t)$, where $x_c(t)$ consists of multiple utterances separated by silence intervals. The mixture is divided into $S$ segments of length $L$ with hop size $H$ ($0 < H \leq L$).
Each segment $y_s(t)$ is processed by a speaker-independent separation model to obtain $C$ outputs $\{\hat{x}_{s,1}(t), \ldots, \hat{x}_{s,C}(t)\}$. 
Since speaker identities are unknown, the assignment of speakers and separated outputs is arbitrary and inconsistent between segments.
The unknown permutation for segment $s$ is denoted as $\pi_s \in \Pi_C$, where $\Pi_C$ denotes the set of all permutations of $\{1, 2, \ldots, C\}$, and $\pi_s(c)$ indicates which speaker the $c$-th separated output corresponds to. 
The cross-segment permutation problem aims to predict the sequence of permutations $\{\pi_1^*, \ldots, \pi_S^*\}$ such that separated segments from the same speaker are consistently assigned to the same output index across all segments.
The number of active speakers and utterances in each segment $y_s(t)$ is unknown and variable.
\vspace{-1ex}
\section{Proposed Method}
\label{sec:proposed-method}
Separation quality varies across segments for long recordings, and segments where speakers are insufficiently separated or where residual noise is prominent easily yield low-quality embeddings. As illustrated in Fig.~\ref{fig-problem}, the proposed method maintains a separate reference pool for each speaker and retains only the most representative embeddings via a TopN selection strategy. Since high cosine similarity scores can only arise from embeddings belonging to the same speaker, the reference pool naturally converges toward high-quality same-speaker embeddings over time while progressively displacing unreliable ones, remaining robust to noise and capable of accumulating representative embeddings across arbitrarily long recordings.

\vspace{-1ex}
\subsection{Quality-Driven Dynamic Clustering}
\label{subsec:dynamic-clustering}
We extract speaker embeddings $\{\mathbf{e}_{s,1}, \ldots, \mathbf{e}_{s,C}\}$ from separated outputs of segment $s$ using a pretrained speaker embedding model, as shown in Fig.~\ref{fig-problem}(b).
For each segment $s$, there are $C$ separate dynamic reference pools $\mathcal{R}_{s,c}$, where each pool contains up to $N$ tuples $(\mathbf{e}_j^{\text{ref}}, q_j)$ of embeddings $\mathbf{e}_j^{\text{ref}}$ and their corresponding quality scores $q_j$ from previously processed segments.
The method assumes at least one active speaker in the first $N$ segments.
This assumption is easily satisfied in practice, as leading silence 
regions can be removed by simple preprocessing before processing.
For the first segment, we use identity permutation where $\pi_1(c) = c$ for all $c$, and the embeddings $\{\mathbf{e}_{1,1}, \ldots, \mathbf{e}_{1,C}\}$ are added to their corresponding reference pools with quality score initialized to $0.5$.
For each segment $s > 1$, the permutation is determined by finding the assignment that maximizes the quality-weighted similarity between current separated output embeddings and reference embeddings in the reference pools,
\begin{equation}
\pi_s^* = \arg\max_{\pi \in \Pi_C} \sum_{c=1}^{C} \sum_{(\mathbf{e}_j^{\text{ref}}, q_j) \in \mathcal{R}_{s-1,c}} q_j \cdot \text{sim}(\mathbf{e}_{s,\pi(c)}, \mathbf{e}_j^{\text{ref}}),
\label{eq:dynamic-clustering}
\end{equation}
where $\text{sim}(\cdot, \cdot)$ denotes cosine similarity, which is widely used for comparing speaker embeddings due to its effectiveness in capturing angular relationships between high-dimensional vectors~\cite{snyder2018x,desplanques2020ecapa}.
The quality scores $q_j$ are used to emphasize representative embeddings over less characteristic ones in Eq.~(\ref{eq:dynamic-clustering}).
\vspace{-1ex}
\subsection{Quality Assessment}
\label{subsec:quality-assessment}
The quality score $q_s$ for segment $s$ is computed as the weighted cosine similarity between all speakers' embeddings and their corresponding reference pools as
\begin{equation}
q_s = \frac{1}{C \cdot |\mathcal{R}_{s-1,c}|} \sum_{c=1}^{C} \sum_{(\mathbf{e}_j^{\text{ref}}, q_j) \in \mathcal{R}_{s-1,c}} q_j \cdot \text{sim}(\mathbf{e}_{s,c}^*, \mathbf{e}_j^{\text{ref}}),
\label{eq:quality-metric}
\end{equation}
where $\mathbf{e}_{s,c}^*$ is the embedding extracted from the separated output assigned to speaker $c$ using predicted $\pi_s^*$ from Eq.~(\ref{eq:dynamic-clustering}). 
\vspace{-1ex}
\subsection{Reference Pool Management}
\label{subsec:pool-management}
As shown in Fig.~\ref{fig-problem}(c), the reference pools are updated by retaining the top $N$ highest-quality embeddings from the union of the current embeddings and existing reference pool embeddings,
\begin{equation}
\mathcal{R}_{s,c} = \begin{cases}
\mathcal{R}_{s-1,c} \cup \{(\mathbf{e}_{s,c}^*, q_{s})\} & \text{if } |\mathcal{R}_{s-1,c}| < N \\
\text{TopN}(\mathcal{R}_{s-1,c} \cup \{(\mathbf{e}_{s,c}^*, q_{s})\}) & \text{otherwise}
\end{cases}
\label{eq:pool-update}
\end{equation}
where $N$ is the maximum pool size, the quality score $q_s$ is from Eq.~(\ref{eq:quality-metric}), and $\text{TopN}(\cdot)$ selects the $N$ highest-score embeddings.
\vspace{-1ex}
\section{Experimental settings}
\label{sec-sxperimental-settings}
\vspace{-1ex}
\subsection{Datasets}
\label{subsec:datasets}
We generate long speech test sets with $2$ to $4$ speakers to evaluate permutation alignment under varying conditions.
The $2$-speaker dense and sparse test sets follow the generation approach in~\cite{wang2025multi}.
For each mixture, we randomly select two speakers from LibriSpeech \textit{train-clean-360}~\cite{panayotov2015librispeech} and concatenate all utterances for each speaker in random order. We use \textit{train-clean-360} rather than \textit{test-clean} due to the limited number of speakers and fewer utterances per speaker in \textit{test-clean}, which are insufficient for generating long speech test sets. Random silence intervals are inserted before each utterance, with durations uniformly sampled from $[1, 10]$ seconds for the dense test set and $[10, 30]$ seconds for the sparse test set. The overlap rate is defined as the ratio of overlapping speech duration to the total duration containing active speech.
Room acoustic simulation and real-world noise addition follow the parameters as~\cite{wang2025multi}. 
All signals are single-channel at $16$~kHz sampling rate with no speaker overlap between training and test sets.
Each test set contains $200$ mixtures. The dense test set achieves overlap rates from $31.2\%$ to $69.7\%$ with an average of $50.1\%$, resulting in a total duration of $121$ hours with mixture lengths averaging $36$ minutes (range: $34$-$38$ minutes). The sparse test set achieves overlap rates from $8.5\%$ to $29.4\%$ with an average of $19.8\%$, resulting in a total duration of $213$ hours with mixture lengths averaging $64$ minutes (range: $54$-$76$ minutes).

We generate $3$-speaker and $4$-speaker sparse long speech test sets following the same simulation procedure and parameter settings as the $2$-speaker case, with at most $3$ speakers active simultaneously, following the setup in~\cite{chen2020continuous}.

\vspace{-1ex}
\subsection{Evaluation Metrics}
\label{subsec:eval}
We evaluate the permutation solutions using SI-SDR~\cite{LeRoux2018a} and segment-wise permutation accuracy. SI-SDR is computed on the entire long speech signal after stitching all segments, reflecting both within-segment separation quality and cross-segment permutation consistency.
We determine oracle permutations for each segment between separated outputs $\hat{x}$ and ground truth reference signals $x^{\text{ref}}$ as
\begin{equation}
\pi_s^{\text{oracle}} = \arg\min_{\pi \in \Pi_C} \sum_{c=1}^{C} \mathcal{L}_{\text{SI-SDR}}(\hat{x}_{s,\pi(c)}, x_{s,c}^{\text{ref}}).
\label{eq:oracle-permutation}
\end{equation}
From the accuracy calculation, we exclude segments where all reference signals are silent using a root mean square (RMS) energy threshold of $-30$~dB, as speaker permutations are undefined in these segments. The segment-wise permutation accuracy is computed as
\begin{equation}
\text{Accuracy} = \frac{1}{|\mathcal{S}_{\text{valid}}|} \sum_{s \in \mathcal{S}_{\text{valid}}} \mathbf{1}[\pi_s^* = \pi_s^{\text{oracle}}],
\label{eq:permutation-accuracy}
\end{equation}
where $\mathcal{S}_{\text{valid}}$ denotes the set of non-silent segments, $\pi_s^*$ is the predicted permutation, and $\mathbf{1}[\cdot]$ is the indicator function that equals $1$ when the condition is true and $0$ otherwise. To account for potential global speaker assignment ambiguity, we evaluate all possible global permutations of speaker assignments and select the one with the highest accuracy.
\vspace{-1ex}
\subsection{Baselines and Pretrained Models}
\label{subsec:baseline}
We compare the proposed method against five baselines. CSS-sep~\cite{chen2020continuous} stitches separated segments by matching overlapping regions between adjacent segments. CSS-sep + VAD + embedding + $k$-clustering~\cite{von2024meeting} applies VAD to CSS outputs, extracts speaker embeddings from speech-active regions, and reassigns output streams via global $k$-means clustering. Sep + speaker inventory~\cite{han2021continuous} identifies single-speaker regions to build a fixed speaker reference, which is established once and unchanged throughout the recording for permutation guidance. Sep + VAD + embedding + UPGMC clustering~\cite{bredin2023pyannote} extracts speaker embeddings from VAD-segmented speech regions of each separated segment and groups them using an agglomerative hierarchical clustering with centroid linkage (UPGMC algorithm).
Sortformer~(AOSC)~\cite{medennikov2025streaming} is an end-to-end online neural diarization system. We use the released model\footnote{\href{https://huggingface.co/nvidia/diar\_streaming\_sortformer\_4spk-v2}{huggingface.co/nvidia/diar\_streaming\_sortformer\_4spk-v2}} and run it on the mixture to obtain global speaker diarization labels, which are then converted to segment-wise permutations for cross-segment alignment of the separated outputs following~\cite{kalda2024pixit}. The model was trained on $2445$ hours of real conversations and $5150$ hours of simulated mixtures including LibriSpeech, which potentially overlaps with the test sets used in this work.
\begin{table}[t]
\centering
\caption{Performance comparison of different systems
using 8\,s segments with 25\% overlap.
Dense scenario: overlap rate 50.1\%, unprocessed mixture SI-SDR $-1.4$\,dB.
Sparse scenario: overlap rate 19.8\%, unprocessed mixture SI-SDR $-1.2$\,dB.
Values show segment-wise permutation accuracy(\%) / SI-SDR(dB).}
\vspace{-1.5ex}
\label{tab:overall}
\resizebox{\columnwidth}{!}{
\begin{tabular}{lcc}
\toprule
System & Dense & Sparse \\
\midrule
Oracle
    & --/11.8  & --/14.3 \\
CSS-sep~\cite{chen2020continuous}
    & 59.1/-3.0  & 53.7/-4.5 \\
CSS-sep + VAD + emb.\ + $k$-clustering~\cite{von2024meeting}
    & 97.1/11.2  & 89.3/8.2 \\
Sep + speaker inventory~\cite{han2021continuous}
    & 85.0/5.7  & 74.1/3.5 \\
Sortformer~(AOSC)~\cite{medennikov2025streaming}
    & 85.2/5.9  & 71.1/2.7 \\
Sep + VAD + emb.\ + UPGMC-clustering~\cite{bredin2023pyannote}
    & 91.3/9.2  & 71.6/2.8 \\
\midrule
Sep + emb.\ + dynamic clustering (proposed)
    & \textbf{98.5/11.7}  & \textbf{96.8/13.3} \\
\bottomrule
\end{tabular}
}
\vspace{-3ex}
\end{table}

All methods use the same FTRNN separator trained following the original implementation~\cite{wang2025multi}, with a $2$-output version for $2$-speaker experiments and a $3$-output version for $3$-speaker and $4$-speaker experiments.
Methods requiring VAD employ a pre-trained pyannote voice activity detection model\footnote{\href{https://huggingface.co/pyannote/voice-activity-detection}{huggingface.co/pyannote/voice-activity-detection}}~\cite{plaquet2023powerset}. For methods requiring speaker embeddings, we employ a pre-trained ECAPA-TDNN model\footnote{\href{https://huggingface.co/speechbrain/spkrec-ecapa-voxceleb}{huggingface.co/speechbrain/spkrec-ecapa-voxceleb}}~\cite{desplanques2020ecapa} trained on VoxCeleb~\cite{nagrani2017voxceleb}. Speaker inventory employs a pre-trained speaker diarization model\footnote{\href{https://huggingface.co/pyannote/speaker-diarization-3.1}{huggingface.co/pyannote/speaker-diarization-3.1}}~\cite{bredin2023pyannote} to identify single-speaker segments. All pretrained models operate in a zero-shot manner, as their training data does not overlap with the separation model's training data. Dynamic clustering uses a reference pool size of $N=10$, where the sensitivity to this parameter is examined in the ablation study. 
\vspace{-1ex}
\section{Results and Discussions}
\label{sec-Experimental-Results}
\vspace{-1ex}
\subsection{Overall Performance}
\label{subsec:overall-performance}
Table~\ref{tab:overall} presents the overall performance of all methods on the $2$-speaker dense and sparse long speech scenarios. 
CSS-sep achieves $59.1\%$ and $53.7\%$ permutation accuracy on the dense and sparse scenarios respectively. 
It is worth noting that CSS is not designed for global permutation alignment but rather aims to minimize overlap in output streams.
CSS-sep + VAD + embedding + $k$-clustering performs substantially better in the dense scenario ($97.1\%$), building on the local utterance continuity provided by CSS. 
In the sparse scenario, performance drops to $89.3\%$, as the CSS stitching stage becomes less reliable under longer utterance gaps and the higher proportion of silence increases the risk of VAD false detections.
Sep + speaker inventory achieves $85.0\%$ and $74.1\%$ on the dense and sparse scenarios, showing consistent but moderate performance in both cases.
Sortformer~(AOSC) achieves $85.2\%$ and $71.1\%$ on the dense and sparse scenarios respectively. Despite potential speaker overlap between its training data and our test sets, the performance remains moderate, suggesting that the generalization of end-to-end diarization systems is sensitive to mismatches in acoustic conditions such as noise, reverberation, and overlap rate between training and test scenarios. 
Sep + VAD + embedding + UPGMC clustering achieves $91.3\%$ in the dense scenario, lower than CSS + VAD + embedding + $k$-clustering. 
In the sparse scenario, performance drops more severely to $71.6\%$, possibly due to increased VAD false detections under high silence proportions, which introduces unreliable embeddings into the clustering process.
\begin{figure*}
\centering
\includegraphics[width=0.99\linewidth]{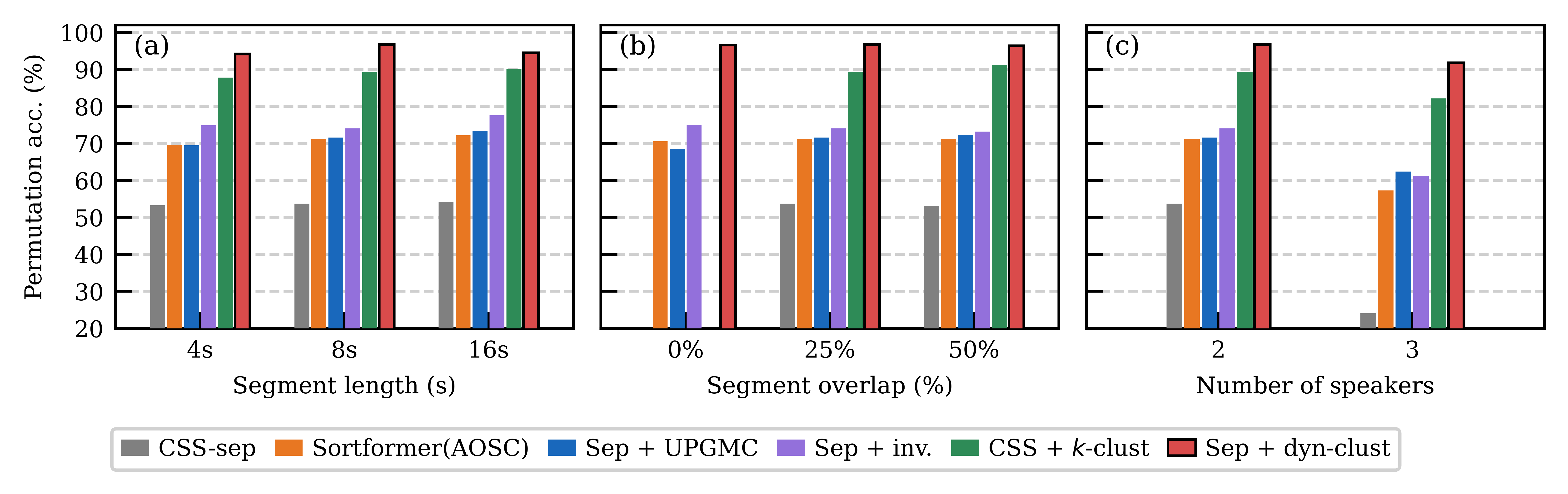}
\vspace{-2ex}
\caption{Permutation accuracy on the sparse long speech scenario as a function of (a) segment length (segment overlap=$25\%$, $2$ speakers), (b) segment overlap (segment length=$8$~s, $2$ speakers), and (c) number of speakers (segment length=$8$~s, segment overlap=$25\%$).}
\label{fig-perm-acc}
\vspace{-2ex}
\end{figure*}

The proposed method achieves $98.5\%$/$11.7$\,dB in the dense scenario and $96.8\%$/$13.3$\,dB in the sparse scenario, respectively.
The SI-SDR of $11.7$\,dB in the dense scenario is very close to the oracle of $11.8$\,dB, indicating that the proposed method approaches the upper bound of achievable performance. The higher oracle SI-SDR in the sparse scenario ($14.3$\,dB vs.\ $11.8$\,dB) reflects that lower overlap rates make within-segment separation inherently easier. Across all methods, the larger performance gap from oracle in the sparse scenario confirms that permutation accuracy is the primary bottleneck.
\vspace{-1ex}
\subsection{Analysis on Sparse Scenario}
\label{subsec:sparse-analysis}
As shown in Fig.~\ref{fig-perm-acc}(a), all methods show relatively stable performance across segment lengths in the 
$2$-speaker sparse scenario, with minor improvements for longer segments. As shown in Fig.~\ref{fig-perm-acc}(b), CSS-sep and CSS-sep + VAD + embedding + $k$-clustering require non-zero overlap and are therefore not applicable at $0\%$ overlap, while the remaining methods show stable performance across overlap rates. 
As shown in Fig.~\ref{fig-perm-acc}(c), all methods experience performance degradation when the number of speakers increases from 2 to 3, reflecting the increased permutation complexity with more speakers. CSS-sep shows the most severe degradation, approaching near-random performance. The proposed method achieves the smallest relative degradation and maintains the highest accuracy among all methods.
For scenarios with more speakers, the separator output count is typically determined by the maximum number of simultaneously active speakers rather than the total speaker count~\cite{chen2020continuous}, and results under such settings are evaluated in Section~\ref{subsec:speaker-count-mismatch}.

\vspace{-1ex}
\subsection{Ablation Study}
\label{subsec:ablation-study}
Table~\ref{tab:ablation-study} presents an ablation study on the $2$-speaker sparse scenario examining the contribution of each component in dynamic clustering. Experiments $1$-$4$ demonstrate the importance of quality weighting mechanisms. Removing quality weighting from permutation decisions (experiment $2$) reduces accuracy by $1.0\%$, while removing quality weighting in pool updates (experiment $3$) causes severe degradation of $6.0\%$. 
The method exhibits robustness to various parameter settings. Reference pool size variations (experiments $5$-$6$) show minimal performance impact, with $N=5$ and $N=2$ achieving comparable results to $N=10$. Initial quality score settings (experiments $7$-$8$) have negligible effect on final performance. 
Similarity metric choice proves critical, as replacing cosine similarity with Euclidean distance (experiment 9) causes severe performance degradation to $68.2\%$ accuracy. 
\begin{table}[t]
\centering
\caption{Ablation study on dynamic clustering components using $8$~s segments with $25\%$ overlap in sparse scenario. Values show segment-wise permutation accuracy(\%)/SI-SDR(dB).}
\vspace{-1ex}
\label{tab:ablation-study}
\resizebox{\columnwidth}{!}{
\begin{tabular}{c|cc|c|c|c|c}
\toprule
\multirow{2}{*}{\#} & Quality Weighting & Quality Weighting & \multirow{2}{*}{Similarity} & \multirow{2}{*}{$N$} & Init. & \multirow{2}{*}{Acc/SI-SDR} \\
& in Eq.~(\ref{eq:dynamic-clustering}) & in Eq.~(\ref{eq:quality-metric}) & & & Score & \\
\midrule
1 & \checkmark & \checkmark & Cosine & 10 & 0.5 & 96.8/13.3 \\
2 & $\times$ & \checkmark & Cosine & 10 & 0.5 & 95.8/12.7 \\
3 & \checkmark & $\times$ & Cosine & 10 & 0.5 & 90.8/10.6 \\
4 & $\times$ & $\times$ & Cosine & 10 & 0.5 & 90.6/10.5 \\
\midrule
5 & \checkmark & \checkmark & Cosine & 5 & 0.5 & 96.3/13.0 \\
6 & \checkmark & \checkmark & Cosine & 2 & 0.5 & 96.3/12.9 \\
\midrule
7 & \checkmark & \checkmark & Cosine & 10 & 0.25 & 96.8/13.3 \\
8 & \checkmark & \checkmark & Cosine & 10 & 0.75 & 96.8/13.3 \\
\midrule
9 & \checkmark & \checkmark & Euclidean & 10 & 0.5 & 68.2/1.5 \\
\bottomrule
\end{tabular}
}
\vspace{-2ex}
\end{table}
\begin{figure}[tb]
\centering
\includegraphics[width=0.95\linewidth]{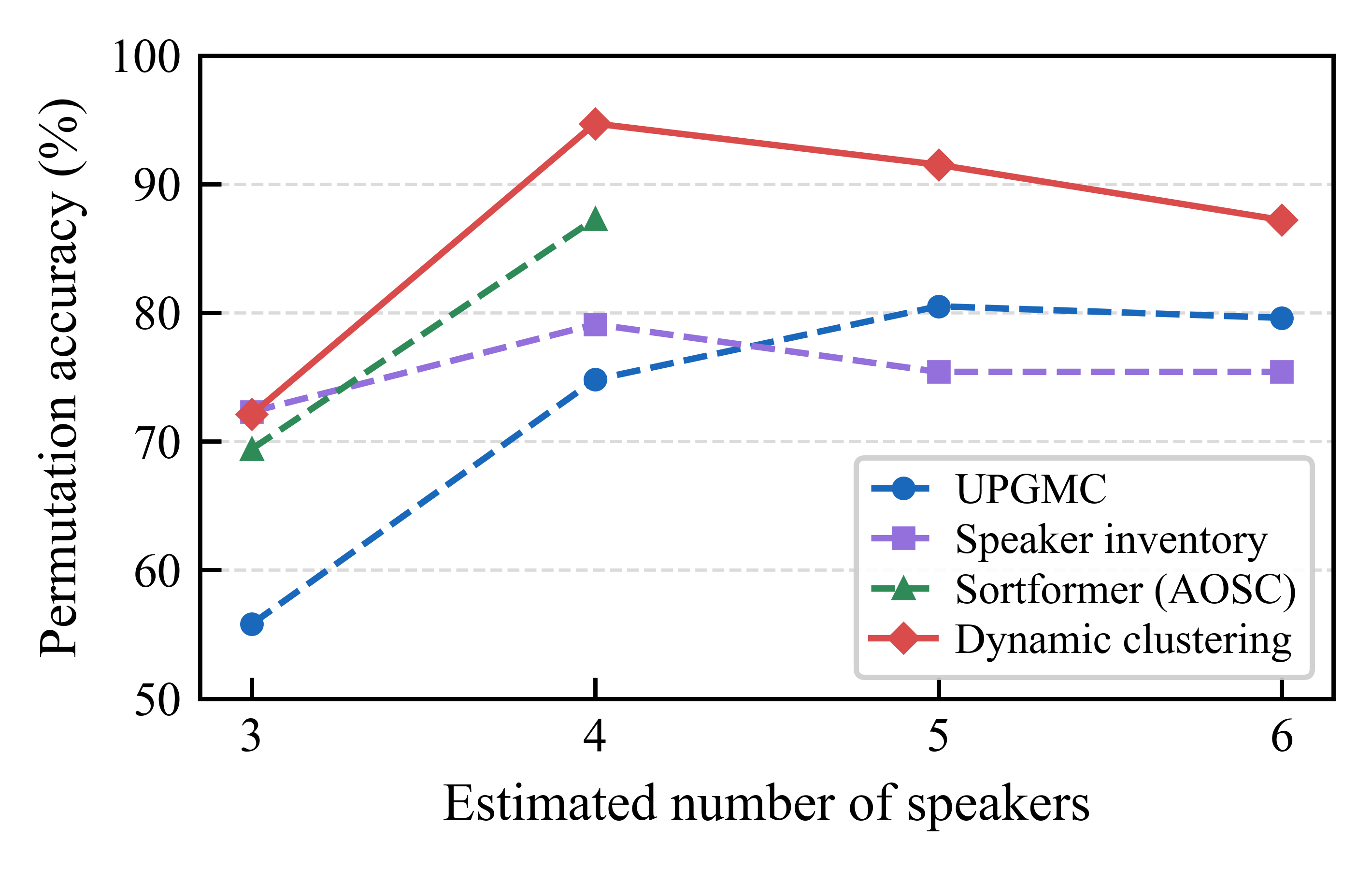}
\vspace{-1ex}
\caption{Permutation accuracy as a function of estimated speaker count $\hat{C}$ on the $4$-speaker sparse long speech test set (segment length=$8$~s, segment overlap=$25\%$).}
\label{fig-speaker-count}
\vspace{-2ex}
\end{figure}
\vspace{-1ex}
\subsection{Robustness to Speaker Count Mismatch}
\label{subsec:speaker-count-mismatch}
All experiments above assume the number of speakers $C$ to be known in advance. In practice, speaker count estimation errors are unavoidable when $C$ is unknown. Although speaker counting is beyond the scope of this work, we analyze the effect of estimation errors by sweeping the estimated speaker count $\hat{C} \in \{3, 4, 5, 6\}$ on the $4$-speaker sparse long speech test set using the $3$-output separator, covering cases where the speaker count is correctly estimated, underestimated, and overestimated.

CSS-based methods are excluded from this comparison, as they aim to minimize overlap in output streams rather than performing global permutation alignment, and are therefore not dependent on $\hat{C}$. For Sortformer (AOSC), we report only $\hat{C}=3$ and $\hat{C}=4$, since the released model supports at most four outputs. For baseline methods, each is run with $C = \hat{C}$ following the same procedure as in the $C$-known experiments.
The proposed dynamic clustering is primarily developed for the case where the number of speakers is known and equals the number of separator outputs. When $\hat{C} > 3$, the assumption no longer holds, and a warm-up initialization stage is introduced to handle this mismatch. 
Embeddings from the first $10$ segments are collected and clustered into $\hat{C}$ groups via $k$-means to initialize the $\hat{C}$ reference pools.
Each segment's $3$ separator outputs are then matched to the reference pools and assigned to $3$ of the $\hat{C}$ output streams, producing $\hat{C}$ complete output streams for the full recording. Since the permutation accuracy in Eq.~(\ref{eq:permutation-accuracy}) assumes $\hat{C} = C$, we adopt a modified permutation accuracy for this experiment 
where a global optimal matching between the $\hat{C}$ estimated streams and the $C$ true speakers is first established across the full recording, and a segment is counted as correct if all its separator outputs are assigned to the matched speakers under this global mapping.

Fig.~\ref{fig-speaker-count} shows the permutation accuracy as a function of $\hat{C}$. Across all methods, underestimation causes larger performance degradation than overestimation, as assigning fewer streams than the true speaker count forces multiple speakers to share the same output, making correct permutation alignment fundamentally impossible for the merged speakers. In contrast, overestimation introduces additional streams, while correct alignment of the existing speakers remains achievable. Unlike other methods whose performance decreases under overestimation, UPGMC clustering shows a consistent performance increase with larger $\hat{C}$, suggesting that additional streams provide more flexibility for separating acoustically similar speaker embeddings during global clustering. Compared to all baselines, the proposed method maintains the highest accuracy across all $\hat{C}$ values, with a peak of $94.7\%$ at the correct estimate $\hat{C} = 4$. This consistent advantage suggests that the quality-weighted mechanism remains effective in scenarios where the number of speakers is unknown, even when the speaker count is incorrectly estimated.
\vspace{-1ex}
\subsection{Embedding-Space Visualization of Dynamic Clustering}
\label{subsec:embedding-visualization}
\begin{figure}[tb]
\centering
\includegraphics[width=0.95\linewidth]{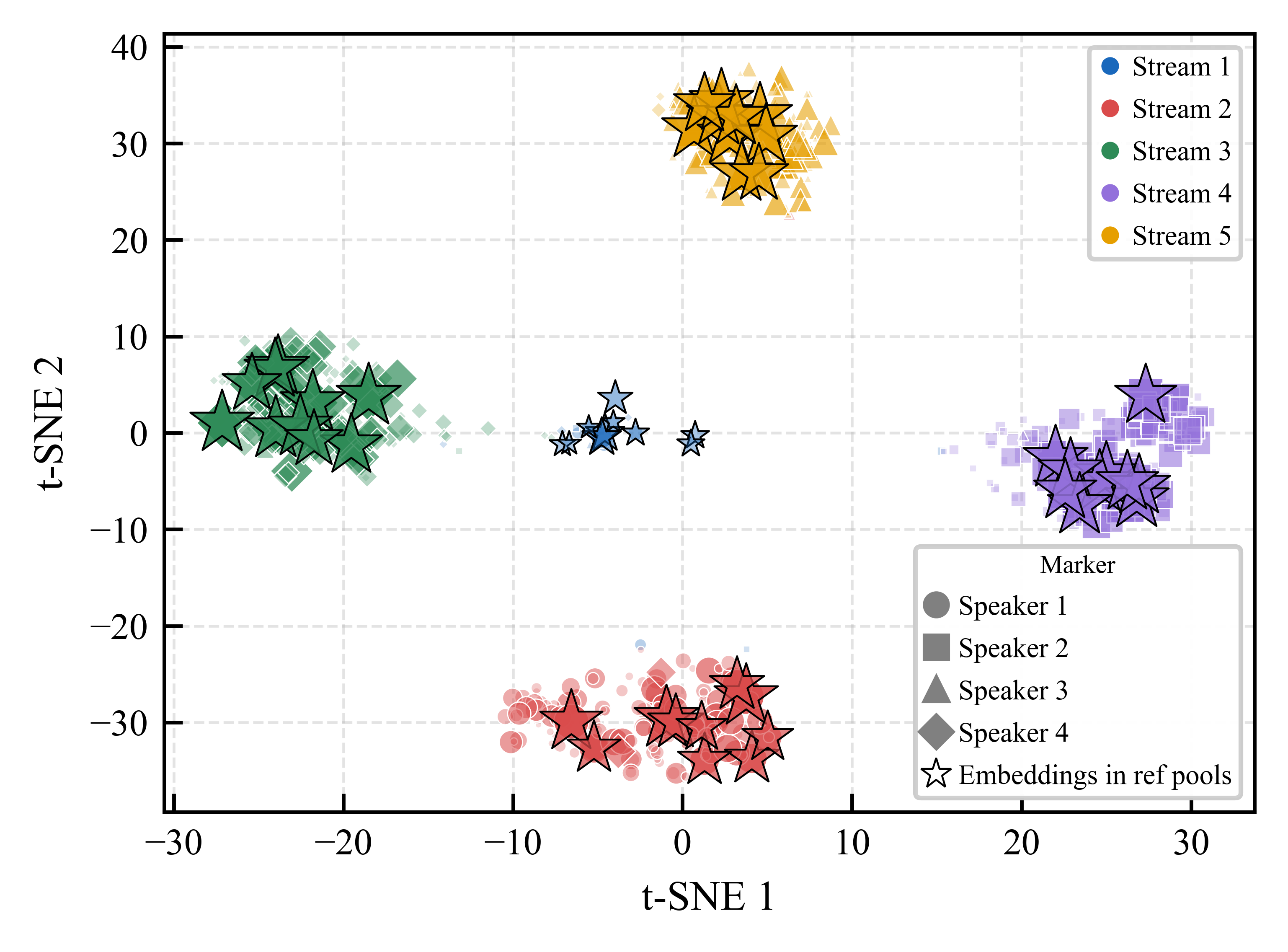}
\vspace{-1ex}
\caption{t-SNE visualization of speaker embeddings after permutation assignment by the proposed dynamic clustering, on a $4$-speaker sparse long speech sample with $\hat{C} = 5$. Colors indicate assigned output streams, marker shapes indicate the true speaker identity, and marker size and opacity reflect the quality score of each embedding. Star markers denote embeddings retained in the reference pools.}
\label{fig-tSNE}
\vspace{-2ex}
\end{figure}

To further examine the behavior of the proposed method under speaker count mismatch, Fig.~\ref{fig-tSNE} visualizes the speaker embedding distributions using t-SNE~\cite{van2008visualizing} for a randomly selected sample from the $4$-speaker test set with $\hat{C} = 5$, following the same experimental setup as Section~\ref{subsec:speaker-count-mismatch}. 

The embeddings form five clusters in the t-SNE space, with each cluster corresponding to one output stream. 
Within each cluster, embeddings from the same true speaker dominate, confirming that the proposed method correctly assigns embeddings to their corresponding output streams.
The cluster corresponding to Stream~1 contains significantly fewer embeddings than the others, with the majority being star markers, suggesting that most embeddings in this cluster were assigned during reference pool initialization with few subsequent additions, which explains the high permutation accuracy of $91.5\%$ achieved 
by the proposed method at $\hat{C} = 5$ in Section~\ref{subsec:speaker-count-mismatch}. Most star markers corresponding to \mbox{Stream~2--5} are concentrated near the center region of each cluster, confirming that the proposed reference pool strategy successfully retains the highly representative embeddings for each speaker even under speaker count mismatch.
\vspace{-1ex}
\section{Conclusions}
\label{sec:conclusions}
This paper addressed the cross-segment permutation problem in long speech separation by proposing a dynamic clustering approach built upon the speaker cache strategy. The method maintains dynamic reference pools of high-quality speaker embeddings and makes permutation decisions based on weighted cosine similarity, requiring no additional training and functioning as a plug-and-play post-processing module compatible with existing separation models. Experimental results confirm that dynamic clustering achieves robust permutation accuracy and maintains speaker consistency across output streams even with extended utterance gaps, which is essential for human listening applications. 
Experiments in the unknown speaker count scenario further demonstrate that the proposed method maintains reliable permutation accuracy and a consistent advantage over baselines even when the speaker count is incorrectly estimated.

\newpage
\bibliographystyle{IEEEbib}
\bibliography{Bib_YWang}

\end{document}